# Thermal Activation Bottleneck in TADF OLEDs based on m-MTDATA:BPhen

Nikolai Bunzmann[1], Douglas L. Baird[2], Hans Malissa[2], Sebastian Weissenseel[1],

Christoph Boehme[2], Vladimir Dyakonov[1] and Andreas Sperlich[1*]

[1]Experimental Physics VI, Julius Maximilian University of Würzburg, 97074 Würzburg, Germany

[2]Department of Physics and Astronomy, University of Utah, Salt Lake City, Utah 84112, USA

(Dated: August 13, 2020)

**Abstract**

Organic light emitting diodes (OLEDs) based on thermally activated delayed fluorescence (TADF) can be highly efficient because of the conversion of non-radiative triplet to radiative singlet states by reverse intersystem crossing (RISC). Even highly efficient TADF OLEDs are limited by long excited state lifetimes though, which limit current densities and cause device degradation. When singlet-triplet energy gaps are comparable to thermal energies (~tens of millielectronvolts), RISC is fast and limited only by bottlenecks due to spin-selection rules. We have studied this phenomenon under device operating conditions using pulsed electrically detected magnetic resonance spectroscopy (pEDMR) in OLEDs based on the donor:acceptor combination m-MTDATA:BPhen (4,4',4"-tris[phenyl(m-tolyl)amino]triphenylamine : 4,7-diphenyl-1,10-phenanthroline). These experiments showed magnetic resonance signatures of emissive exciplex states at the donor:acceptor interface, yet these signals did not reveal coherent spin propagation effects. Instead, the intensity of these signals scales linearly with the energy dose of the applied microwave pulses. This observation excludes the direct involvement of the resonantly prepared coherent spin states and indicates that the observed current response is due to magnetic resonant heating. This implies that the studied TADF blend is not limited by spin-forbidden RISC, but rather by the thermal activation step.

*sperlich@physik.uni-wuerzburg.de

## I. Introduction

Organic light emitting diodes (OLEDs) that that are able to utilize thermally activated delayed fluorescence (TADF) require the upconversion of non-radiative triplet states to radiative singlet states via reverse intersystem crossing (RISC) (Goushi2012, Uoyama2012). One approach to realize TADF OLEDs is to use proper combinations of donor and acceptor molecules between which emissive exciplex states form. Due to the extent of such intermolecular excited states, their wavefunction overlap is small, resulting in small singlet-triplet energy gaps $\Delta E_{ST}$ that enable RISC at ambient temperatures. While emission from singlet excitons occurs on the order of nanoseconds, TADF is a slow process with time constants in the range of microseconds (Goushi2012, Uoyama2012). This causes issues that hinder the widespread application of TADF OLEDs. Long excited state lifetimes enhance the probability of second-order processes, such as triplet-triplet and triplet-polaron annihilation, which limit the device efficiency at current densities relevant for display applications (efficiency roll-off) and ultimately lead to device degradation (Murawski2013, Grüne2020). It is unclear what the actual rate-limiting process is: the thermal upconversion or the spin-forbidden triplet-to-singlet RISC. The latter depends directly on the initial spin state and, thus, the coherent propagation of triplet exciton spin states, should be reflected in transient RISC rates. In order to investigate this upconversion mechanism, a method with high spin sensitivity and selectivity is required. These requirements are met by methods based on electron paramagnetic resonance (EPR), which is a widely used experimental tool for studies where artificially induced spin propagation modulates spin-dependent physical observables (Weil2007). In the context of TADF, only few EPR studies have been reported (Ogiwara2015, Väth2017, Evans2018). Recently, we demonstrated the application of photoluminescence, electroluminescence and electrically detected magnetic resonance (PLDMR, ELDMR, EDMR) on donor:acceptor based TADF OLEDs (Bunzmann2020a, Bunzmann2020b). We were able to determine that the intermediate spin states that are involved in the light generation of the investigated TADF materials are different for optical and electrical excitation. To further address the question of whether spin-coherent processes are involved in TADF, a method is required that allows for the observation of spin-dynamics. Continuous wave (cw) magnetic resonance experiments have only a limited capability to provide such information as different electronic and spin-transition processes such as exciton lifetimes as well as spin-relaxation times can affect magnetic resonance spectra in a similar fashion, implying large ambiguities in the interpretation of the measurements (Schnegg2012). This problem can be remediated by dynamic time-domain magnetic resonance spectroscopy such as pulsed EDMR (pEDMR) where the evolution of spin-dependent electronic transition rates is monitored while spin states are coherently driven by magnetic resonant microwave (MW) radiation pulses, allowing direct observation of the relationship between the probed spin species and the optoelectronic properties of the studies devices (Boehme2003a, Schnegg2012, Boehme2017). PEDMR has previously been applied to a variety of materials and devices, including silicon-based bulk and thin film devices (Boehme2003b, Boehme2004, Herring2009), organic solar cells (Behrends2010, Kupijai2015), as well as polymer-based OLEDs (McCamey2008, McCamey2010, Baker2011, Baker2015), revealing comprehensive information about spin dependent recombination and transport mechanisms in these systems. PEDMR applied to first-generation polymer-based singlet emitter OLEDs (that are limited by spin-dependent recombination), showed that coherent spin manipulation of charge carriers in the OLED is reflected by observable current and electroluminescence (EL) changes (McCamey2008, Baker2012, Kavand2016). This experimental approach is used here for the study of electronic transitions in TADF OLEDs in order to elucidate the influence of spin-selection rules on RISC in donor:acceptor interface devices.

**II. Results**

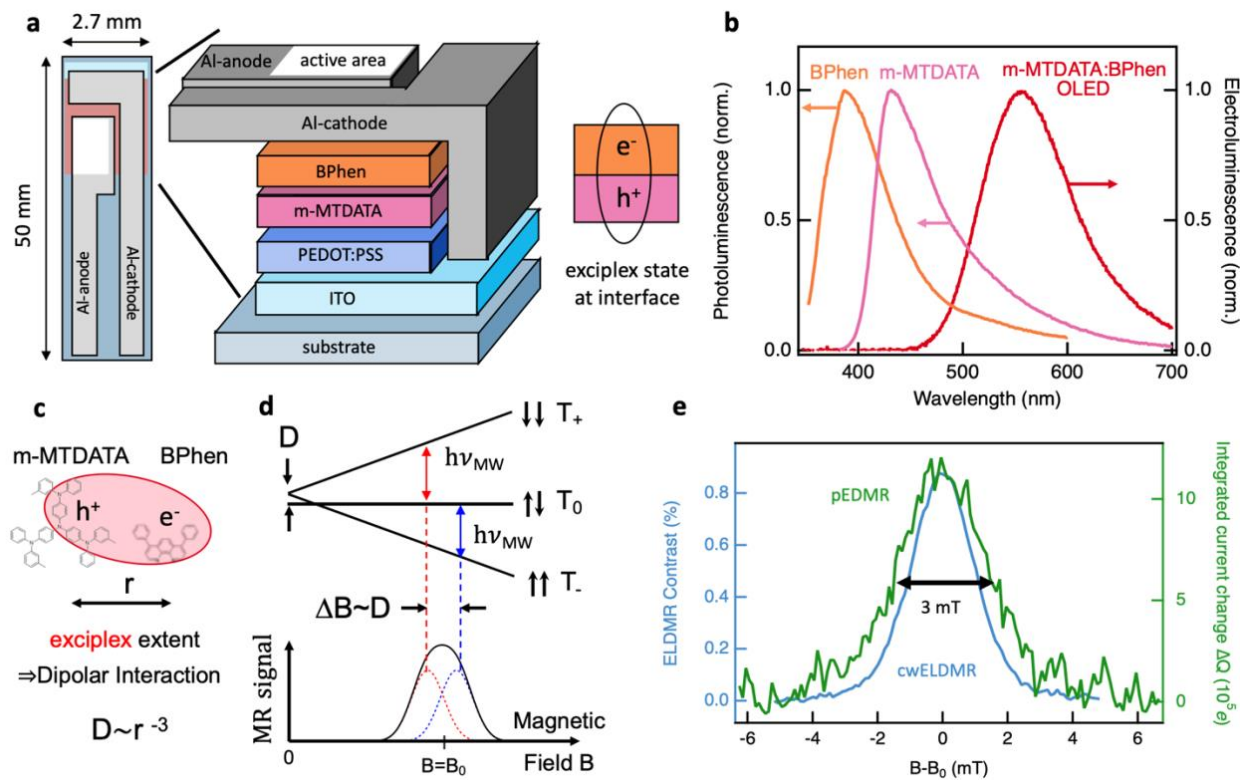

**Figure 1.** (a) Top and side view of the OLED layer stack. (b) EL spectrum of an OLED based on m-MTDATA:BPhen and PL spectra of the pristine donor and acceptor materials, respectively. (c) Illustration of a delocalized exciplex state between m-MTDATA and BPhen molecules. (d) Zeeman splitting of a spin triplet in a magnetic field and magnetic resonance signal due to overlapping microwave induced transitions between triplet sublevels. (e) Comparison of cwELDMR and pEDMR OLED spectra measured at RT. The magnetic field axis is shifted, such that resonance peaks at $B = B_0$ are centered around $B - B_0$. Both methods yield a Gaussian shaped signal which is assigned to the exciplex triplet $^3$Exc.

We investigated exciplex-based TADF OLEDs using 4,4',4"-tris[phenyl(m-tolyl)amino]triphenylamine (m-MTDATA) as a donor and 4,7-diphenyl-1,10-phenanthroline (BPhen) as an acceptor. Delayed fluorescence in this system becomes apparent from time-resolved photoluminescence (trPL) measurements on m-MTDATA:BPhen mixed solid films where a delayed component in the µs range can be identified. This delayed component was previously assigned to thermally activated RISC (Zhang2014, Bunzmann2020a). The formation of an exciplex state enables a small singlet-triplet gap $\Delta E_{ST}$ of a few tens of millielectronvolts (Zhang2014, Bunzmann2020a) and, therefore, efficient triplet harvesting. The used sample architecture and layer stack are shown in **Figure 1a**. The sample geometry is optimized for magnetic resonance measurements in an X-band EPR setup which requires the width of samples to be limited to 2.7 mm. The active area (marked as a white rectangle in Figure 1a) in this layout is defined by the overlap of indium tin oxide (ITO) and the aluminum-cathode which is around 5 mm2. In this area, the layer stack is given by ITO (120 nm) / PEDOT:PSS (40 nm) / m-MTDATA (30 nm) / BPhen (40 nm) / Ca (5 nm) / Al (120 nm). Poly(3,4-ethylenedioxythiophene) polystyrene sulfonate (PEDOT:PSS) is spin-coated from an aqueous solution and m-MTDATA from a chlorobenzene solution. BPhen, Ca and Al were subsequently evaporated on top. Electrons injected from the cathode and holes injected from the anode form exciplex states at the interface of m-MTDATA and BPhen. The formation of these states is apparent from a comparison of photoluminescence (PL) spectra from pristine m-MTDATA and BPhen films and an EL spectrum from an OLED that is processed as described above. As shown in **Figure 1b**, a clear redshift between the PL spectra and the EL spectrum occurs, demonstrating the formation of exciplex states. The emission layer in such types of OLEDs is therefore given by the two-dimensional interface between m-MTDATA and BPhen.

Exciplex triplets are $S = 1$ states and their mutual splitting is modulated by an external magnetic field $B$. If microwaves of matching energy $h\nu_{\text{MW}}$ are applied, transitions between these states become possible when the resonance condition

$$h\nu_{\text{MW}} = g\mu_B B \Delta m_s \pm D\left(\cos^2\theta - \tfrac{1}{3}\right)$$

is fulfilled. Here, $h$ represents Plank's constant, $g$ the Landé-factor of the exciplex, $\mu_B$ the Bohr magneton, $\Delta m_s$ the allowed change of spin quantum number, $D$ the dipolar interaction between the two charge carriers forming the triplet state, and $\theta$ the angle between the vector of the external magnetic field and the vector connecting the two spins of the triplet state. The magnetic field dependent splitting is illustrated in **Figure 1d,** where the energetic triplet sublevels $T_+$, $T_0$, and $T_-$ are plotted as a function of an external magnetic field. As shown, the energetic degeneracy of those three sublevels is absent even at zero magnetic field, due to the magnetic dipolar interaction $D$ of the two electronic states which form the exciplex triplet system. The energy of the dipolar splitting scales with $r^{-3}$, where $r$ is the distance between the two charge carrier states. For exciplex triplets this distance is relatively large (≳1 nm) (Bunzmann2020a), as the electron is located on the acceptor molecule and the hole on the donor molecule (**Figure 1c**). In magnetic resonance experiments, $\nu_{\text{MW}}$ is typically fixed while the magnetic field $B$ is swept across resonance. Because of the dipolar interaction $D$, two resonant transitions occur at different magnetic field values. However, if $D$ is sufficiently small in comparison to the linewidth and the exciplex orientation angle $\theta$ is random, the separate resonance peaks are not resolved and single featureless signals with Gaussian to Voigt-type lineshapes appear, whose widths $\Delta B$ are directly dependent on $D$.

As discussed in detail elsewhere (Bunzmann2020a), signals originating from exciplex triplets 3Exc can be observed with spin-sensitive optical detection of PL, as well as by EL, or through current detection from OLEDs under magnetic resonance conditions. In all these experiments, the relative changes of PL, EL or device current are recorded at constant driving voltage, while manipulating the spin manifolds which control these observables. A cwELDMR spectrum for a m-MTDATA:BPhen OLED is shown in **Figure 1e**. The almost identical pEDMR spectrum demonstrates that time-resolved, pulsed magnetic resonance can be used in order to probe the same exciplex spin species. The nature of this signal under different excitation regimes will be discussed in the following.

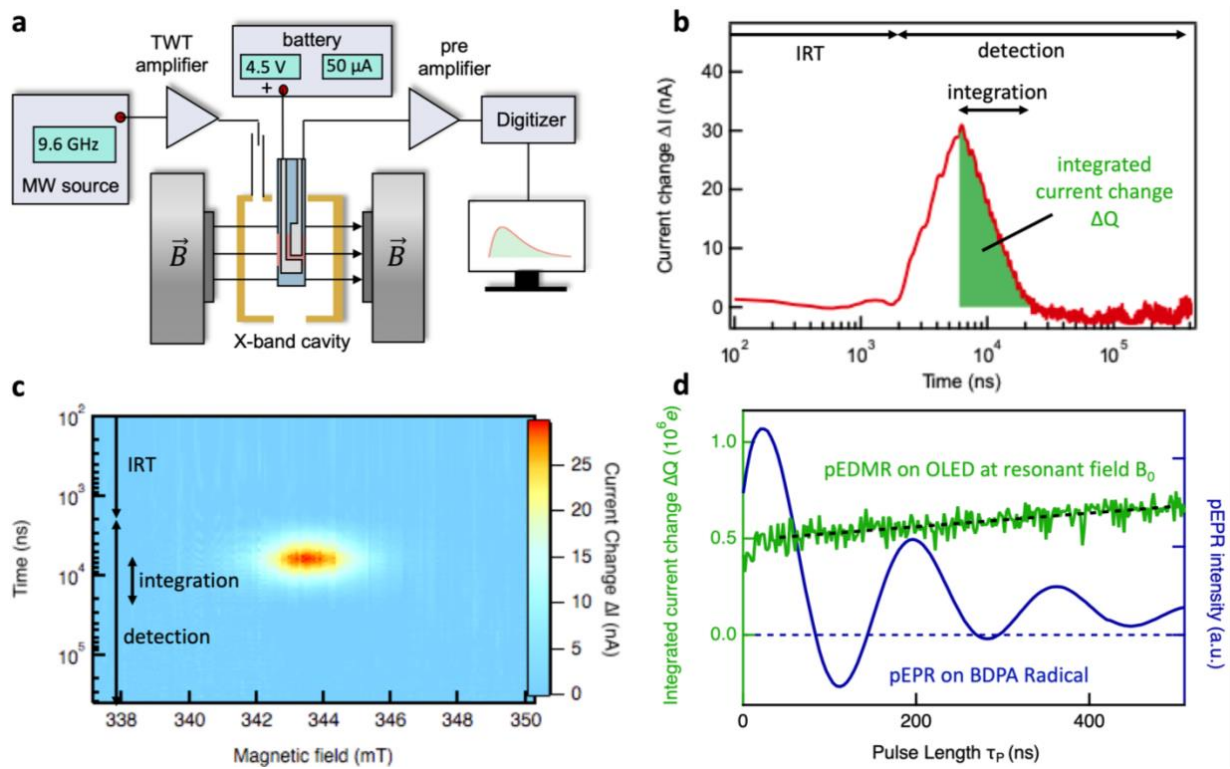

**Figure 2.** **(a)** Basic pEDMR setup. For details see text. **(b)** Current-change of an m-MTDATA:BPhen OLED as a function of time after a $\tau_P = 4\,\mu s$ MW radiation pulse that is in magnetic resonance with the exciplex states. Transient recording starts with the rise of the MW pulse (X-band, ~9.6 GHz). The instrument response time (IRT) delays signal detection for around 2 μs. An integration window of 16 μs starts 2 μs after the MW pulse ends, yielding the integrated current change $\Delta Q$ **(c)** 2D plot of current change transients similar to the data shown in (b) as a function of magnetic field, from below magnetic resonance to above magnetic resonance values. **(d)** Blue trace: pEPR of a BDPA spin reference sample demonstrating the ability of the applied radiation pulses to induce coherent spin motions, i.e. Rabi oscillation. Green trace: $\Delta Q$ as a function of the MW pulse length $\tau_P$ at resonant magnetic field $B_0$ for an OLED under identical conditions as for the pEPR experiments. The black dashed line is a guide for the eye. A linear signal rise instead of Rabi oscillations can be observed.

**Figure 2a** shows the basic components of the pEDMR setup used for these experiments. It is based on a Bruker E580 EPR spectrometer. An OLED is inserted into an X-band dielectric cavity (Bruker ER 4118 X-MD5) that is placed inside an electromagnet that creates the external magnetic field $B$. MW pulses are generated by the E580 MW bridge with an adjustable attenuator that allows to vary the pulse power. The pulses are amplified by a traveling wave tube (TWT) amplifier with maximum output power of 1 kW before being directed to the cavity. The MW frequency of the pulses is resonant with the dimensions of the cavity and in the range of 9.6 GHz. The OLED is electrically driven by a rechargeable isolated voltage source (Stanford Research SIM928) at 4.5 V and a current density of 1 mA/cm$^2$. MW induced changes of the current $\Delta I$ through the OLED are pre-amplified by a current-voltage transimpedance amplifier (Stanford Research SR570) and detected by the built-in transient recorder of the E580. During a pEDMR experiment, MW pulses of defined length and power are applied to the OLED while it is biased with a constant voltage that drives a device current $I$. Resonant microwave radiation induces transitions between triplet sublevels as described in **Figure 1c**. The coherent propagation of triplet populations will then affect recombination and transport mechanisms causing a time-dependent current change $\Delta I$. (Lepine1972, Kaplan1978, Kishimoto1981) For all measurements, the origin of the time axis coincides with the rise of the MW pulse.

**Figures 2b** and **2c** display the results of pEDMR experiments using a pulse length of $\tau_P = 4\,\mu s$ and a power attenuation of 9 dB. The observation of a current change is delayed for 2 µs after the MW pulse begins, owing to the instrument response time (IRT). Subsequently, the signal rises and reaches a maximum at around $t \sim 6\,\mu s$. The signal has decayed to zero (steady state) at ~30 µs. PEDMR measurements are frequently superimposed by non-magnetic resonant high-frequency background signal artifacts (see **Appendix A** for details), that can be due to induction effects of the OLED wiring. These are, therefore, not related to spin-dependent processes in exciplex triplets and were subtracted for clarity. **Figure 2c** shows a 2D plot of current change transients in dependence of the external magnetic field measured under the same conditions as in **Figure 2b**. In order to quantify the entirety of the observable magnetic resonance-induced number of exciplex transitions, it is convenient to integrate $\Delta I$ over time, yielding a charge $\Delta Q$ (Boehme2003a). To make $\Delta Q$ comparable for different pulse lengths, an integration window of 16 µs was chosen, starting 2 µs after the end of the MW pulse. Those time limits reasonably take the IRT and signal lifetime of each transient into account. Determining $\Delta Q$ for each transient yields the pEDMR spectrum shown in **Figure 1e**. The shape of this curve is identical to the exciplex signal detected by cwELDMR except for a slightly broader linewidth caused by MW power broadening (Waters2015, Jamali2017) that becomes detectable under the intensive pulsed drive conditions needed for pEDMR.

Before coherent pEDMR experiments can be performed, proper calibration of MW pulse length and power is essential. To adjust the pulses with respect to the spin flip time between the involved Zeeman sublevels (see **Figure 1c**), we used pulsed EPR on a well-known spin reference sample, the radical α,γ-Bisdiphenylene-β-phenylallyl (BDPA) (Eaton2010, Kavand2017). A tiny (nanomole) amount of BDPA was mounted on top of the OLED to ensure identical experimental conditions for both pEPR and pEDMR. The observed Rabi oscillations of the BDPA spin-½ system are shown in **Figure 2d** for 9 dB MW attenuation. The fitting of this trace to an exponentially decaying sine wave allows the estimation of the duration for a π-pulse to 85 ns and the microwave magnetic field $B_1 = 0.21$ mT. The oscillation decays with a time constant of 144 ns that is limited by the spectrometer and the spin-spin coherence time $T_2^*$, which is in line with literature values (Raedt2012).

Following the calibration of the spectrometer and demonstrating that proper conditions for pEDMR have been established for the coherent excitation of exciplex spin-triplets in OLEDs under operating conditions, all prerequisites for the straight-forward detection of exciplex spin-Rabi oscillations are met. For this experiment, the integrated current change $\Delta Q$ at resonance is recorded as a function of the applied MW pulse length $\tau_P$. Such experiments have previously been performed on singlet emitter OLEDs based on various π-conjugated polymers, e.g. derivatives of the polymer poly(p-phenylene vinylene) (PPV) using this particular pEDMR setup (McCamey2008, McCamey2010, Baker2011). Remarkably, the pEDMR $\Delta Q$ trace in **Figure 2d** does not contain any sign of Rabi oscillations. Instead, after an initial instrument-related offset, $\Delta Q$ follows a linearly rising trend; i.e. the detected charge $\Delta Q$ appears proportional to the energy dose of the spin-resonantly absorbed MW pulse. To further explore this effect and corroborate that this is not an experimental artefact, we performed a series of experiments with varying pulse length and MW power as shown in **Figure 3a,b**.

First, we confirmed both, that the non-resonant offset as well as the Voigt-shaped exciplex signal increase monotonically with increasing pulse energy (power × length). The linear increase of the non-resonant offset (**Figures 3e**) can be explained by sample heating due to MW fields. Charge transport in organic semiconductor devices is a temperature-activated hopping transport. The observation of a linear current increase for an OLED operated at constant voltage is thus to be expected for Joule heating of the device.

The increase of the resonance signal with pulse energy is more complex as MW power broadening is involved (Eaton2010). At low power, the signal has the expected hyperfine field and powder distributed zero-field splitting induced Gaussian lineshape as evident by the cwELDMR spectrum plotted in **Figure 1e**, that was recorded with approximately an order of magnitude lower MW magnetic field strength $B_1$ than the lowest-power pEDMR spectra. At high MW power, the shape becomes Lorentzian. Voigt fits show this transition from Gaussian to Lorentzian lineshape (**Figure 3d**), as well as a doubling of the FWHM linewidth (**Figure 3c**) for the tested MW pulse energy range (Jamali2017). The signal peak amplitude is plotted in **Figure 3f** demonstrating signal increase up to ~1 mJ microwave energy and then no further amplitude increase. This observation indicates saturation of the EPR transitions between the exciplex triplet sublevels and is consistent with the power broadening (Waters2015, Jamali2017).

The observation that signal parameters in **Figures 3** directly depend on pulse energy, irrespectively of whether pulse power or length were changed, demonstrates that the observed pEDMR current changes do not reflect coherent spin motion, and thus do not display Rabi oscillations as suggested by the data in **Figure 2d**.

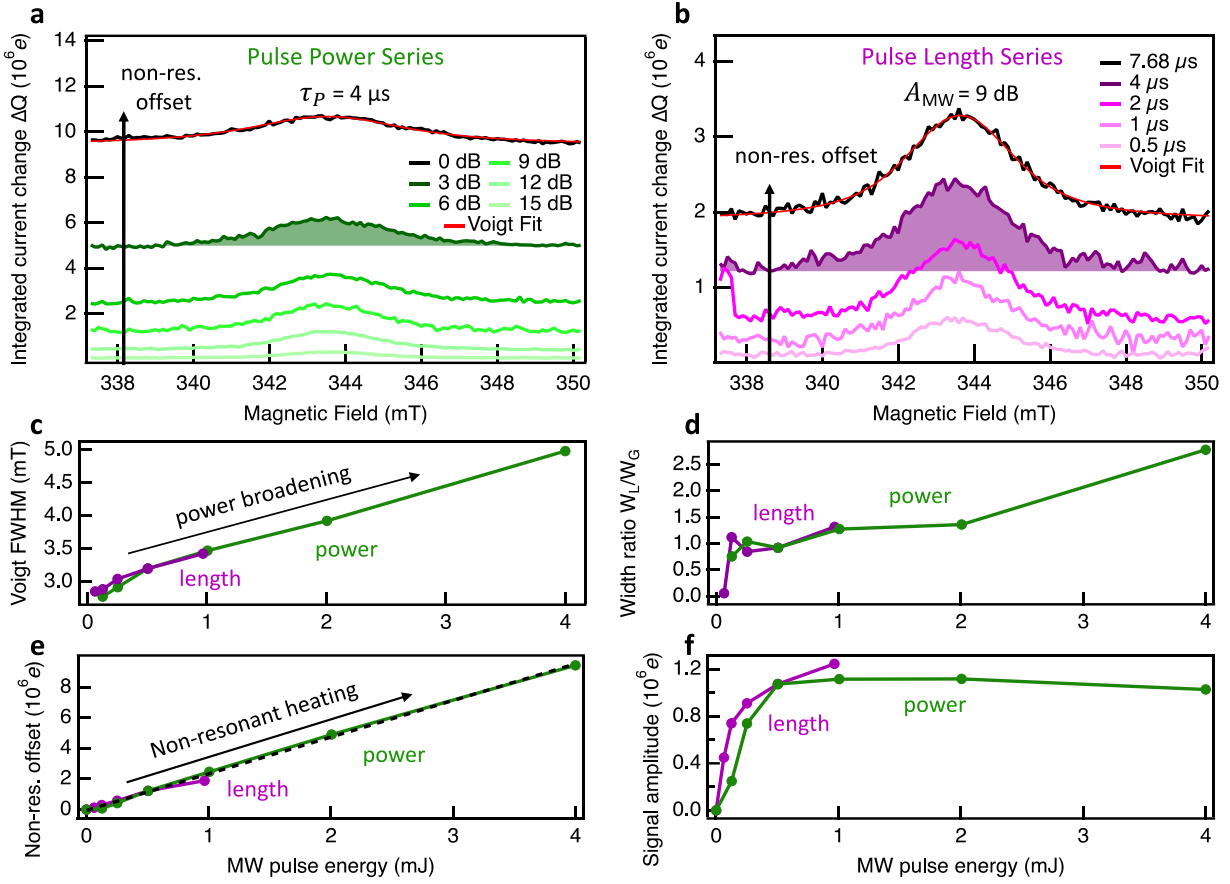

**Figure 3.** (a) $\Delta Q$ as a function of the magnetic field $B$ for different MW pulse attenuations (pulse length fixed at $\tau_P = 4$ μs) and (b) for different pulse lengths (attenuation $A_{MW} = 9$ dB). All spectra are fitted with a Voigt line shape (red). The fit parameters are plotted versus the MW pulse energy (product of pulse power and duration). (c) Voigt linewidth FWHM increase representing power broadening. (d) Ratio of the width of the Lorentzian contribution $W_L$ and the Gaussian contribution $W_G$. (e) Non-resonant offset that is caused by Joule heating of the OLED. (f) On-resonance signal amplitude. Dashed black traces represent guides to the eye. The signals depend directly on pulse energy, irrespective of whether pulse length or power are modified.

## III. Discussion

Our observations of magnetic resonant and non-magnetic resonant heating as well as the absence of detectable coherent spin-motion effects can be understood in the context of the generally accepted model of TADF OLEDs (Uoyama2012, Goushi2012, Bunzmann2020a) that is illustrated in **Figure 4**. Singlet and triplet excitons localized on separate donor and acceptor molecules are not involved in these processes as their manifolds are energetically much higher (Huang2014, Zhu2015) than the exciplex states, which are the only states accessible for an electrically driven OLED. Losses by non-radiative decay from ₁Exc and ₃Exc states are omitted for simplicity. First, free charges form exciplex states at the donor:acceptor interface with a statistical 75:25 triplet to singlet ratio, neglecting thermal polarization effects of free charge carriers from these excitations. Singlet exciplexes ₁Exc decay via prompt fluorescence (PF) with sub-microsecond rate constants $k_F$. Triplet exciplexes ₃Exc from a low vibronic excited state $v_0$ are thermally excited into a higher vibronic state $v_i$ with rate $k_{TA}$ (Penfold2018). If the higher vibronic state is approximately isoenergetic with ₁Exc, RISC takes place with rate $k_{RISC}^0$ resulting in delayed fluorescence (DF) that can last over several microseconds.

In an external magnetic field $B$, ₃Exc states energetically split into Zeeman sublevels $T_+$, $T_0$, and $T_-$. Resonant MW pulses drive transitions between these sublevels, leading to a coherent oscillation, demonstrated by the Rabi oscillation of BDPA radicals in **Figure 2d**. Spin lattice relaxation $k_{SLR}$ drives the triplet population back to thermal equilibrium and causes decoherence on a timescale of hundreds of nanoseconds. The following slow thermal activation and RISC ultimately prevents coherent spin manipulation between triplet sublevels from affecting the observed rate of DF and the pEDMR current. Any coherent spin manipulation is averaged by slow TADF processes, similar to a built-in low-pass filter. Even zero IRT would thus not change this outcome. As a consequence, resonant microwaves only cause non-coherent Joule heating of the OLED which lowers the materials' electrical resistance and is observed as a current increase in the pEDMR response. The temperature increase of the OLED can actually be estimated to be <0.1 K by taking pEDMR amplitudes and temperature-dependent current-voltage characteristics into account (see **Appendix B**).

We want to point out that the described spin resonance experiments presented here were executed on an actual 2D interface between donor and acceptor materials. This is a rare opportunity to study spins confined to a 2D system, which usually inhibits the use of any standard EPR methods due to the low overall spin number. The superior sensitivity of EDMR is the key to this kind of experimental approach.

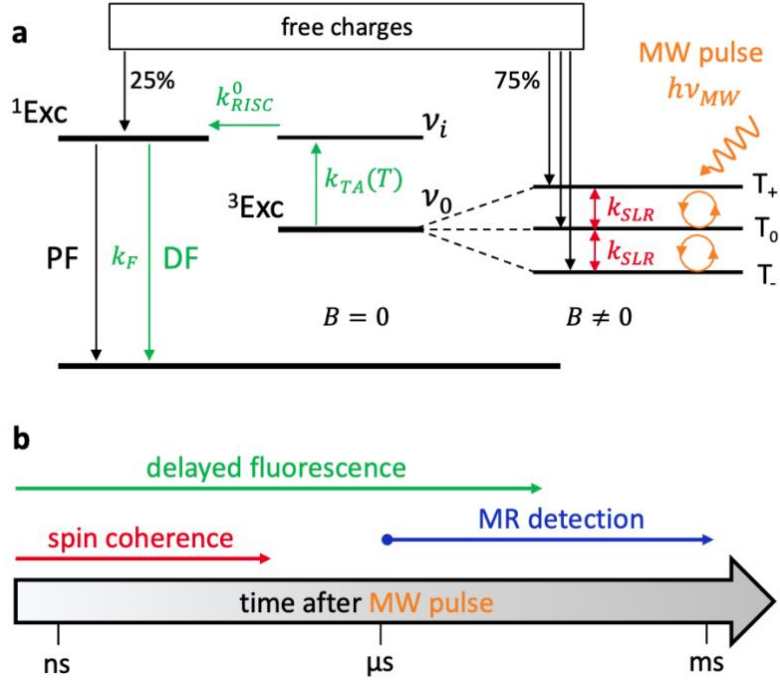

**Figure 4.** Spin coherence and TADF time constants in comparison. **(a)** Energy diagram illustrating the steps involved in the TADF process. Free charge carriers form triplet and singlet exciplex states (3Exc, 1Exc) at the donor:acceptor interface. Singlets recombine radiatively emitting prompt fluorescence (PF) with rate $k_F$. 3Exc triplets are thermally activated ($k_{TA}$) to a higher vibronic state ($v_0 \rightarrow v_i$) that is isoenergetic with 1Exc to enable efficient RISC, followed by delayed fluorescence (DF). In an external magnetic field ($B \neq 0$), 3Exc Zeeman sublevels ($T_+$, $T_0$, $T_-$) are energetically split. Between them, resonant MW pulses $h\nu_{MW}$ drive coherent population oscillations that are dampened by spin lattice relaxation $k_{SLR}$. **(b)** Depiction of involved time scales. DF is observed over several microseconds, while magnetic resonance is detected from ~2 µs onwards. Coherent spin manipulation and SLR take place on the timescale of hundreds of nanoseconds. TADF limiting $k_{TA}$ and $k_{RISC}^0$ time constants lead to equalization of 3Exc spin polarization by $k_{SLR}$.

## IV. Conclusion

This study has been motivated by the question of whether the RISC-rate limiting process in TADF donor:acceptor OLEDs is thermal activation or spin-forbidden conversion of non-emissive triplet to emissive singlet exciplex states. To investigate this problem, we applied pEDMR to OLED devices under typical device operating conditions which showed that device currents are controlled by exciplex triplet states, however, under coherent spin-drive conditions, Rabi oscillations are not observed. This observation, together with the recombination rate dynamics that are discussed above, we conclude that the time constant of thermally activated RISC processes is too long for coherent spin propagation to be reflected by the delayed fluorescence rates. When RISC transitions are slower than spin lattice relaxation transitions, the spin ensemble will thermalize before the thermally activated electronic transition will take place and the spin-excitation will not modulate the RISC rate. The comparison of the dependence of the pEDMR signal magnitude on MW pulse length and MW pulse power, indicates that it is solely related to the energy dose of the pulse, i.e. the total energy that is deposited per MW pulse. We therefore conclude that the observed signals are induced by resonant heating: microwave photons absorbed by triplet exciplexes or by the OLED device itself lead to a slight temperature increase of <0.1 K that results in a detectable change in current and electroluminescence intensity. Thus, our results reveal that RISC in donor:acceptor TADF OLEDs based on mMTDATA:BPhen is indeed not influenced by coherent spin effects because delayed fluorescence takes place on time scales much longer than the involved spin coherence times. Future designs of TADF materials need to take this potential bottleneck into account. We note that this main conclusion of our study may not be universally applicable to all TADF-based OLED systems: More efficient TADF emitters with shorter excited state lifetimes might represent a dynamic regime, where spin effects become rate limiting under OLED operation conditions and, therefore, coherent spin propagation may be detectable by observation of spin-dependent transition rates.

**Acknowledgements**

N.B. and A.S. acknowledge support by the German Research Foundation, DFG, within the SPP 1601 (SP1563/1-1). D.L.B., H.M., and C.B. acknowledge support by the US Department of Energy, Office of Basic Energy Sciences, Division of Materials Sciences and Engineering under Award #DE-SC0000909. S.W. acknowledges DFG FOR 1809 (DY18/12-2). A.S. and V.D. acknowledge the German Research Foundation, DFG, within GRK 2112. We thank Dr. Michael Auth for help with device preparation, as well as Dr. Marzieh Kavand for help with pEDMR measurements.

**Author contributions**

A.S., H.M., V.D. and C.B. designed the experiments. N.B. and S.W. prepared the devices. N.B., S.W. and A.S. carried out electro-optical device characterization. A.S., D.L.B. and H.M. performed the magnetic resonance measurements. A.S., C.B. and V.D. supervised the research project. N.B. and A.S., evaluated and processes the data. N.B. and A.S. drafted the manuscript, which was then iterated by all other authors.

**Competing Financial Interests statement**

The authors declare no competing financial interests.

**Appendix A:** Data processing of pEDMR raw data of an m-MTDATA:BPhen OLED.

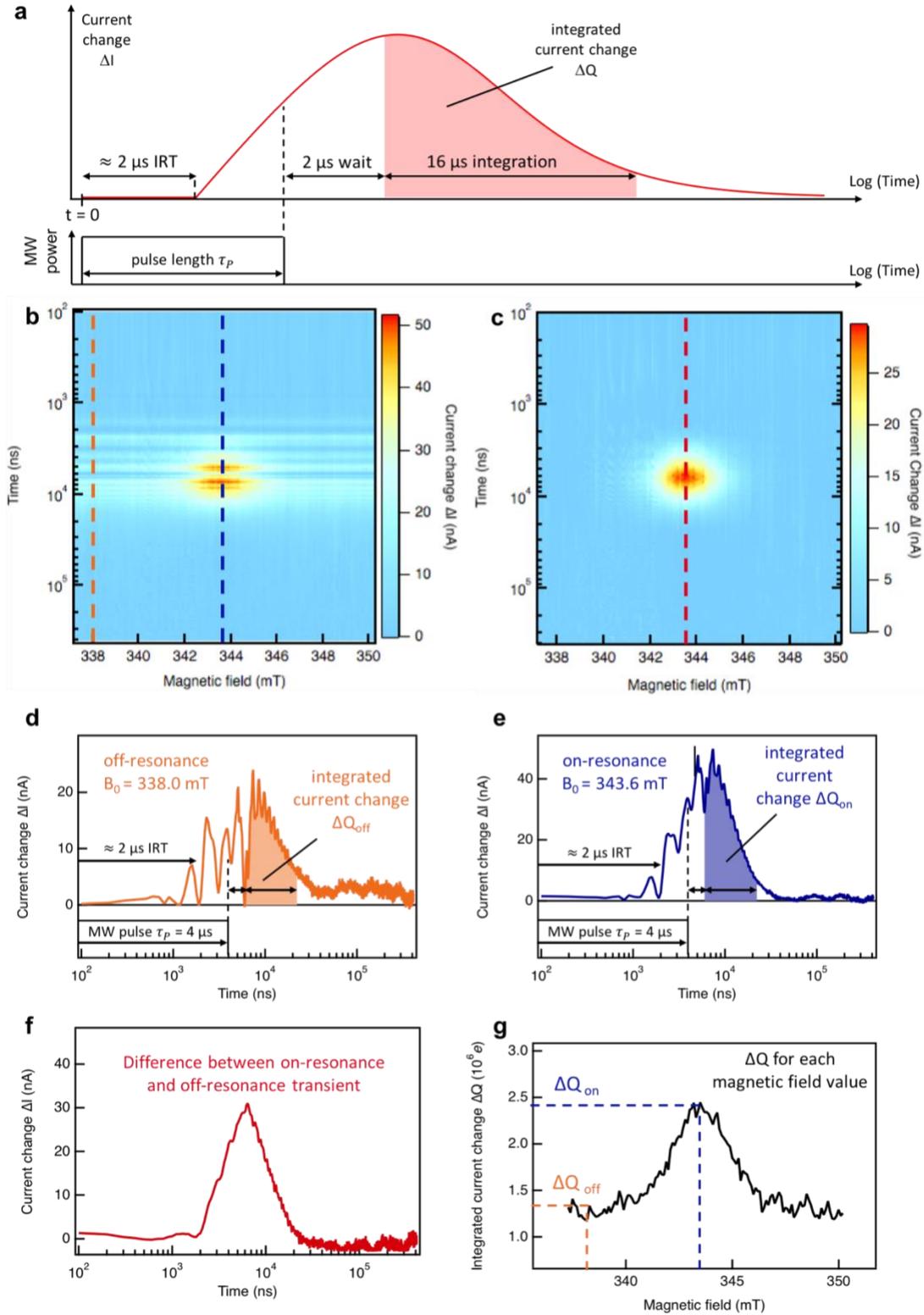

**Figure 5. (a)** Illustration of the detection and evaluation scheme in pEDMR experiments: A MW pulse of length $\tau_P$ is applied to the OLED while the resulting current change $\Delta I$ is recorded. The rise of the MW pulse triggers the start

of data acquisition ($t = 0$). The instrument response time (IRT) delays the signal rise by around 2 µs. In general, the start of a detectable signal and the duration of the MW pulse can overlap. In order to make the integrated current change $\Delta Q$ comparable for different pulse lengths, the integration was chosen as a 16 µs window starting 2 µs after the end of the MW pulse. **(b)** Raw data 2D map of $\Delta I$ transients within the magnetic field range. The MW pulse length is $\tau_P = 4$ µs and the MW power attenuation $A = 9$ dB. Dashed vertical lines mark the off-resonance transient at $B_0 = 338.0$ mT (orange) and the on-resonance transient at $B_0 = 343.6$ mT (blue) that are plotted in **(d,e)**. **(c)** Same 2D map with subtracted off-resonance transient. The high-frequency background oscillations caused by induction in the OLED circuit that are observed for pEDMR (d) can be eliminated this way. The dashed vertical line (red) marks the transient of the current change induced by magnetic resonance and is shown in **(f)** as the difference between on-resonance and off-resonant transients (also shown in Figure 2b). **(g)** Integrated current change $\Delta Q$ within the used magnetic field range. Every data point corresponds to the integrated time trace at the respective magnetic field.

**Appendix B:** Determination of OLED temperature increase by resonant heating

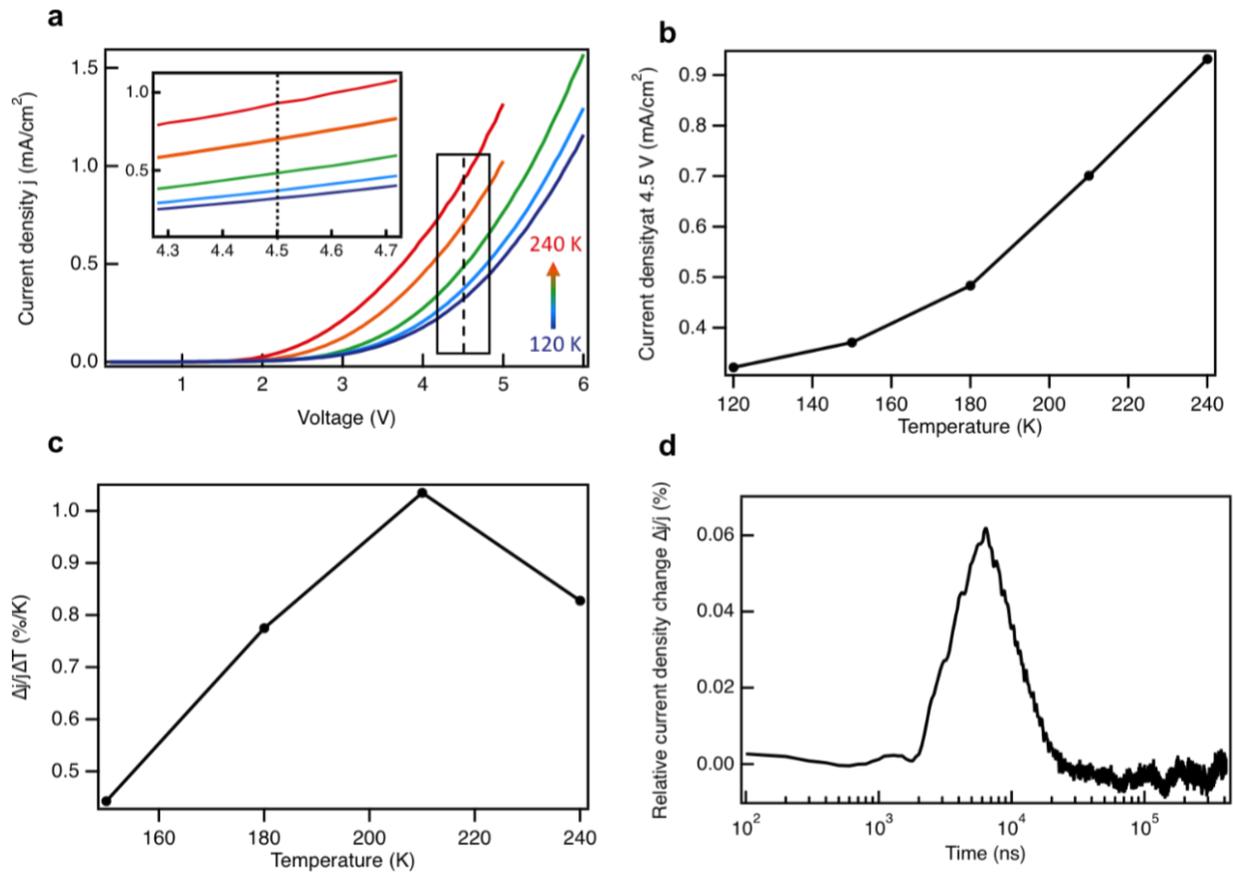

**Figure 6.** **(a)** Temperature dependent current density $j$ versus voltage $V$ characteristics for an OLED based on m-MTDATA:BPhen. Inset: Enlarged section for determination of j at 4.5 V. **(b)** Current density $j$ at 4.5 V for each temperature. **(c)** $\Delta j/(j\Delta T)$ calculated with the data points from b) according to: $\frac{\Delta j}{j\,\Delta T}(T_i) = \frac{j_{i+1}-j_i}{j_i \cdot (T_{i+1}-T_i)}$. A temperature change $\Delta T$ of 1 K results in a relative current change of 0.5-1%. **(d)** Calibrated, relative pEDMR current density change transient $\Delta j(t)/j$. The maximum relative change is 0.06%, which corresponds to an estimated temperature change $\Delta T$ of the OLED by 0.06 K. (Assuming that a temperature change of 1 K changes the current density by 1%.)